\documentclass[twocolumn,prd,preprintnumbers,amsmath,amssymb]{revtex4}

\usepackage{graphicx}
\usepackage{dcolumn}
\usepackage{bm}
\usepackage{epsfig}
\usepackage{amssymb}
\usepackage{amsmath}

\def\scri#1{{\cal I}_#1}

 \def\tskip{\setlength{\tskip}{5pt}}
\def\colwidth{\setlength{\colwidth}{3.5in}}

\def\apjl{Astrophys. J. Lett.}

\def\mnras{Mon. Not. Roy. Astron. Soc.}
\def\apjs{Astrophys. J., Suppl. Ser}

\newcommand{\lsim}{\mathrel{\hbox{\rlap{\lower.55ex\hbox{$\sim$}} \kern-.3em \raise.4ex \hbox{$<$}}}}
\newcommand{\gsim}{\mathrel{\hbox{\rlap{\lower.55ex\hbox{$\sim$}} \kern-.3em \raise.4ex \hbox{$>$}}}}
\newcommand{\beq}{\begin{equation}}
\newcommand{\eeq}{\end{equation}}
\newcommand{\beqa}{\begin{eqnarray}}
\newcommand{\eeqa}{\end{eqnarray}}
\newcommand{\drm}{\mathrm{d}}

\newcommand{\omegaM}{\Omega_\mathrm{M}}
\newcommand{\omegaL}{\Omega_\Lambda}
\newcommand{\omegaX}{\Omega_X}
\newcommand{\lcdm}{$\Lambda$CDM }
\newcommand{\adec}{a_\mathrm{dec}}
\newcommand{\xdec}{x_\mathrm{dec}}
\newcommand{\vxdec}{\vec{x}_\mathrm{dec}}
\newcommand{\zdec}{z_\mathrm{dec}}
\newcommand{\zeq}{z_\mathrm{eq}}
\newcommand{\taudec}{\tau_\mathrm{dec}}
\newcommand{\half}{\frac{1}{2}}
\newcommand{\gradPsi}{\vec{\nabla}\Psi}
\newcommand{\Psip}{\Psi_\mathrm{p}}
\newcommand{\Psikp}{\Psi_\mathrm{p}}

\begin{document}

\title{Superhorizon Perturbations and the Cosmic Microwave Background}

\author{Adrienne L. Erickcek, Sean M. Carroll, and Marc Kamionkowski}
\affiliation{Theoretical Astrophysics, California Institute of Technology, Mail Code
103-33, Pasadena, CA 91125}

\date{\today}

\begin{abstract}
Superhorizon perturbations induce large-scale temperature anisotropies in the cosmic microwave background (CMB) via the Grishchuk-Zel'dovich effect.  We analyze the CMB temperature anisotropies generated by a single-mode adiabatic superhorizon perturbation.  We show that an adiabatic superhorizon perturbation in a \lcdm universe does not generate a CMB temperature dipole, and we derive constraints to the amplitude and wavelength of a superhorizon potential perturbation from measurements of the CMB quadrupole and octupole.  We also consider constraints to a superhorizon fluctuation in the curvaton field, which was recently proposed as a source of the hemispherical power asymmetry in the CMB.
\end{abstract}
   
\maketitle

\section{Introduction}
The finite age of the Universe implies the existence of a cosmological particle horizon beyond which we cannot observe.  Inhomogeneities with wavelengths longer than the horizon are not completely invisible, however.  The generation of large-scale temperature fluctuations in the cosmic microwave background (CMB) by superhorizon perturbations is known as the Grishchuk-Zel'dovich effect \cite{GZ78}.  Through this effect, measurements of the low-multipole moments of the CMB \cite{COBE, WMAP1} place constraints on the amplitudes and wavelengths of superhorizon perturbations.  

A well-known application of the Grishchuk-Zel'dovich effect uses CMB observations to place a lower bound on the size of the nearly homogeneous patch that contains the observable Universe. This bound was first derived for an Einstein-de Sitter universe \cite{GZ78, Turner91}, and then for an open universe \cite{KTF94, GLLW95}.  Most recently, an analysis of the WMAP first-year data \cite{WMAP1} found that our nearly homogeneous patch of the Universe extends to 3900 times the cosmological horizon \cite{CDF03}.  All of these analyses considered a statistically isotropic distribution of power in superhorizon perturbations and then asked how large the wavelength of order-unity perturbations needed to be in order to be consistent with the observed CMB anisotropies.

In this paper, we analyze the CMB anisotropies induced by a single superhorizon adiabatic perturbation mode rather than an isotropic distribution of superhorizon inhomogeneities.   A single-mode superhorizon perturbation to the gravitational potential would naively be expected to generate a dipolar CMB anisotropy with an amplitude comparable to the perturbation amplitude across the observable Universe.  This is not the case in an Einstein-de Sitter universe, however, because the intrinsic dipole in the CMB produced by the perturbation is exactly cancelled by the Doppler dipole induced by our peculiar motion \cite{GZ78, Turner91, BL94}.  We show that the same cancellation occurs for an adiabatic superhorizon perturbation in a flat universe with a cosmological constant ($\Lambda$), cold dark matter (CDM), and radiation.  The strongest constraints to the amplitude and wavelength of a single superhorizon mode therefore arise from measurements of the CMB quadrupole and octupole.  These constraints are less stringent than those derived for modes in a realization of a random-phase random field because it is possible to choose the phase of a single sinusoidal perturbation in such a way that there is no resulting quadrupole anisotropy. 

Single-mode superhorizon perturbations have received attention recently \cite{GHHC05, Gordon07, DDR07, EKC08} because they introduce a special direction in our Universe and could be responsible for observed deviations from statistical isotropy in the CMB \cite{TOH03, OTZH04, LM05, Eriksen04, HBG04, Eriksen07, GE08}.    In particular, we investigated in a recent paper \cite{EKC08} how a superhorizon perturbation during slow-roll inflation can generate an anomalous feature of the CMB: the fluctuation amplitude on large scales ($\ell \lsim 40$) is 10\% larger on one side of the sky than on the other side \cite{Eriksen04, HBG04, Eriksen07}.  We first considered a perturbation to the inflaton field, but we found that the perturbation required to generate the observed power asymmetry induces large-scale anisotropies in the CMB that are too large to be consistent with measurements of the CMB octupole.  We then considered a multi-field model of inflation in which a subdominant field, called the curvaton, is responsible for generating primordial perturbations \cite{Mollerach90, LM97, LW02, MT01}.   We found that a superhorizon perturbation in the curvaton field can generate the observed power asymmetry without inducing prohibitively large CMB anisotropies.  We will use Ref.~\cite{EKC08} as an example of how one may apply the CMB constraints to single-mode superhorizon perturbations derived here.   

We begin in Section \ref{sec:basics} by reviewing the Grishchuk-Zel'dovich effect for adiabatic perturbations.  In Section \ref{sec:potential}, we derive the CMB anisotropy induced by a sinusoidal superhorizon perturbation in the gravitational potential, as would arise from a sinusoidal inflaton fluctuation.   We also show in Section \ref{sec:potential} that a superhorizon adiabatic perturbation does not generate a large dipolar anisotropy in a \lcdm universe because the leading-order intrinsic dipole anisotropy is cancelled by the anisotropy induced by the Doppler effect.  A sinusoidal curvaton fluctuation generates a potential perturbation that is not sinusoidal, and we derive the constraints to single-mode perturbations to the curvaton field in Section \ref{sec:curvaton}.  We summarize our results in Section \ref{sec:conclusions}.    Finally, an analytic demonstration of the dipole cancellation in a \lcdm universe is presented in Appendix \ref{app:cancel}, and the cancellation is shown to occur in flat universes containing a single fluid with an arbitrary constant equation of state in Appendix \ref{app:cancel2}.  

\section{The Grishchuk-Zel'dovich effect: A brief review}
\label{sec:basics}
Working in conformal Newtonian gauge, we take the perturbed Friedmann-Robertson-Walker (FRW) metric to be
\beq
\drm s^2 = -(1+2\Psi)\drm t^2 + a^2(t)\delta_{ij}(1-2\Phi)\drm x^i \drm x^j,
\label{metric}
\eeq
where $a$ is normalized to equal one today.  In the absence of anisotropic stress, $\Psi=\Phi$.  The primary sources of anisotropic stress are the quadrupole moments of the photon and neutrino distributions.  Since the perturbations we consider are superhorizon, the distance travelled by photons and neutrinos arriving at a point is far smaller than the wavelength of the perturbation.  Therefore, the quadrupole moments of the photon and neutrino distributions are much smaller than the monopole moments, and the anisotropic stress is negligible \cite{HS95, HS95b}.  We will assume that $\Psi=\Phi$ throughout this paper.

On large scales, intrinsic fluctuations in the CMB temperature are generated by metric perturbations through the Sachs-Wolfe effect \cite{SW67}.  The current temperature fluctuation at a particular point in the sky (specified by $\hat{n}$) is given by
\beqa
\left[\frac{\Delta T}{T}(\hat{n})\right]_\mathrm{SW+ISW} &=& \frac{\Delta T}{T}(\taudec, \hat{n}\xdec)+\Psi(\taudec, \hat{n}\xdec) \nonumber \\
&&+ 2\int_{\taudec}^{\tau_0} \frac{\drm \Psi}{\drm \tau}[\tau, \hat{n}(\tau_0-\tau)] \drm \tau,
\label{SWandISW}
\eeqa
where $\xdec$ is the comoving distance to the surface of last scattering, and $\tau = \int{dt/a}$ is the conformal time: $\taudec$ is the conformal time at decoupling and $\tau_0$ is the current conformal time.    Given that the early universe was radiation-dominated, the Boltzmann and Einstein equations imply
\beq
 \frac{\Delta T}{T}(\taudec)+\Psi(\taudec) = \Psi(\taudec)\left[2-\frac{5}{3}\left\{\frac{\frac{9}{10}\Psip}{\Psi(\taudec)}\right\}\right],
 \label{SW}
 \eeq
where $\Psip$ is the primordial value of $\Psi$ at $a=\tau=0$.  This expression simplifies to the familiar $\Psi(\taudec)/3$ in the limit that the Universe was matter-dominated at the time of decoupling.  We will refer to this as the Sachs-Wolfe (SW) effect, and the last term in Eq.~(\ref{SWandISW}) will be referred to as the integrated Sachs-Wolfe (ISW) effect.

We also observe a temperature fluctuation due to our peculiar motion \cite{PW68, KK03}, which we will refer to as the Doppler effect:
\beq
\left[\frac{\Delta T}{T}(\hat{n})\right]_\mathrm{D} =  \hat{n}\cdot\vec{v}_\mathrm{net}(\hat{n})+v_\mathrm{net}^2\left[(\hat{n}\cdot\hat{v}_\mathrm{net} )^2-\frac{1}{2}\right] + {\cal O}(v_\mathrm{net}^3),
\label{doppler}
\eeq
where $\vec{v}_\mathrm{net}(\hat{n})$ is our current velocity relative to the fluid at the surface of last scattering in a given direction.  If $\vec{v}(\tau,\vec{x})$ is the proper peculiar velocity of an observer at conformal time $\tau$ and position $\vec{x}$ in the frame defined by Eq.~(\ref{metric}), then
\beq
\vec{v}_\mathrm{net}(\hat{n}) = \vec{v}(\tau_0, \vec{0})-\vec{v}(\taudec, \hat{n}\xdec).
\eeq
For superhorizon perturbations, there is a direct relationship between the potential perturbation $\Psi$ and the proper peculiar velocity of an observer falling into the potential well:
\beq
\vec{v}(\tau,\vec{x}) = -\frac{2a^2}{H_0 \omegaM}\frac{H(a)}{H_0}\left(\frac{y}{4+3y}\right)\left[\gradPsi+\frac{\drm}{\drm \ln a} \gradPsi\right].
\label{veqn}
\eeq
Throughout this paper, $y\equiv a(1+\zeq)$, where $\zeq$ is the redshift of matter-radiation equality, $H(a) \equiv (1/a)\drm a/\drm t$, and $\omegaM$ is the present-day ratio of the matter density to the critical density.  

Thus we see that the SW, ISW, and Doppler effects on the CMB are all determined completely by the evolution of the gravitational potential $\Psi$.   The rest of this Section is devoted to the derivation of $\Psi_{\vec{k}}(a)$ for a single superhorizon adiabatic perturbation mode ($k\ll H_0$) in a flat \lcdm Universe that includes radiation.   In this case, the Hubble parameter is given by
\beq
H^2(a) = H_0^2\left[\frac{\omegaM}{a^4}\left(\frac{1}{1+\zeq}+a\right)+\omegaL\right]
\label{Heqn}
\eeq
with $\omegaL = 1-\omegaM-\omegaM(1+\zeq)^{-1}$.

Since no causal processes can separate the components of the density perturbation, the superhozion perturbation may be treated as a perturbation to a single fluid with over-density in the fluid's rest frame  $\Delta_{\vec{k}}$ and peculiar velocity $\vec{v}_{\vec{k}}$.  These two quantities are related through two coupled equations \cite{HS95}: in a flat Universe with no entropy perturbations these equations are
\beqa
\dot{\Delta}_{\vec{k}}- 3 w aH \Delta_{\vec{k}} &=& -(1+w)kv_{\vec{k}}, \label{vdelta1} \\
\dot{v}_{\vec{k}} + aHv_{\vec{k}} &=& \frac{4}{3}\frac{w}{(1+w)^2} k \Delta_{\vec{k}} + k\Psi_{\vec{k}}, \label{vdelta2}
\eeqa
where an overdot denotes differentiation with respect to $\tau$, and $w \equiv p/\rho$ is the equation of state parameter for the perturbed fluid.  Since matter and radiation are the only perturbed density components, $w = 1/[3(1+y)]$.  

Differentiating Eq.~(\ref{vdelta1}) with respect to $\tau$ gives an expression for $\dot{v}_{\vec{k}}$ that may be used, along with Eq.~(\ref{vdelta1}) itself, to eliminate $v_{\vec{k}}$ from Eq.~(\ref{vdelta2}).   
If matter and radiation are the only perturbed density components, the potential $\Psi$ is related to $\Delta_{\vec{k}}$ through \cite{HS95}
\beq
\Psi_{\vec{k}} = -\frac{3}{4} \left(\frac{k_\mathrm{eq}}{k}\right)^2 \frac{1+y}{y^2} \Delta_{\vec{k}}
\label{PsiDelta}
\eeq
where $k_\mathrm{eq} = (1+\zeq)^{-1}H_\mathrm{eq}$.   This expression may be inserted into Eq.~(\ref{vdelta1}) to obtain Eq.~(\ref{veqn}) for $\vec{v}$, and it may be used to eliminate $\Psi_{\vec{k}}$ from Eq.~(\ref{vdelta2}).  The resulting differential equation for $\Delta_{\vec{k}}$ is
\beqa
&&\ddot{\Delta}_{\vec{k}}+ \frac{y(5+3y)}{(4+3y)(1+y)}aH \dot{\Delta}_{\vec{k}} \label{deltaeqn}\\
&+&\left[\frac{-\dot{H}}{H^2a(1+y)}-\frac{8+3y}{(4+3y)(1+y)}\right]a^2H^2\Delta_{\vec{k}} \nonumber \\
&+&\frac{H_0^2}{4+3y}\left[\frac{4}{3} \frac{k^2}{H_0^2}- \frac{(4+3y)^2}{4y^2}\frac{H_\mathrm{eq}^2}{(1+\zeq)^2H_0^2}\right]\Delta_{\vec{k}} = 0, \nonumber
\eeqa
where $H_\mathrm{eq}^2 \simeq 2\omegaM(1+\zeq)^3H_0^2$ is the Hubble parameter at the time of matter-radiation equality.   Thus we see that the last term on the last line is always much larger than one, 
while $k/H_0 \ll1$ for a superhorizon mode.   Therefore, we will neglect the $k^2$ term in this equation.  

\begin{figure}
\includegraphics[width=8.5cm]{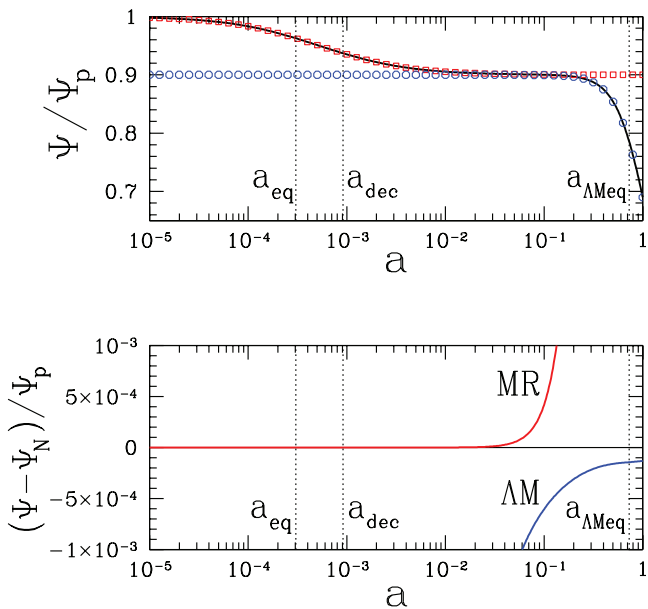}
\caption{The evolution of $\Psi$ in a \lcdm Universe with radiation.  In the top panel, the solid curve is the numerical solution to Eq.~(\ref{Psieqn}), the squares are the analytic solution for matter and radiation (MR) only, and the circles are the solution for $\Lambda$ and matter ($\Lambda$M) only.  The bottom panel shows the difference between these two approximate solutions and the numerical solution.}
\label{fig:Psi}
\end{figure}

Inserting Eq.~(\ref{PsiDelta}) into Eq.~(\ref{deltaeqn}) yields a differential equation for $\Psi$.  It will be more convenient to work with $\Psi(a)$ than $\Psi(\tau)$:
\beqa
&&\Psi^{\prime\prime}_{\vec{k}}(a) + \frac{1}{a}\left(2+\frac{16+9y}{4+3y}+\frac{\drm \ln H}{\drm \ln a}\right)\Psi^\prime_{\vec{k}}(a) \label{Psieqn} \\
&+& \frac{1}{a^2}\left(3+\frac{4}{4+3y}-\frac{4+3y}{4y^4}\frac{H_\mathrm{eq}^2}{H^2}+\frac{\drm \ln H}{\drm \ln a}\right)\Psi_{\vec{k}} = 0, \nonumber
\eeqa
where a prime denotes differentiation with respect to $a$.   Long before $\Lambda$-domination ($\omegaM/a^3 \gg \omegaL$), this equation yields the usual expression for $\Psi$ in a Universe with only matter and radiation \cite{KS84}:
\beq
\Psi_\mathrm{MR}(y) = \frac{\Psip}{10y^3}\left(16\sqrt{1+y}+9y^3+2y^2-8y-16\right).
\label{mrPsi}
\eeq
Long after matter-radiation equality $(y \gg 1)$, Eq.~(\ref{Psieqn}) is solved by 
\beq
\Psi_\mathrm{\Lambda M}(a) = \left[\frac{\frac{9}{10}\Psip}{a}\right]\frac{5}{2} \omegaM H_0^2 \tilde{H}(a) \int_0^a \frac{\drm a^\prime}{[\tilde{H}(a^\prime)a^\prime]^3}
\label{lmPsi}
\eeq
where $\tilde{H}(a)$ is the radiation-free ($\zeq \rightarrow \infty$) limit of $H(a)$.

We numerically solve Eq.~(\ref{Psieqn}) to obtain $\Psi(a)$, which is shown in Fig.~\ref{fig:Psi}.   We use WMAP5+BAO+SN \cite{WMAP5} best-fit values $\omegaM=0.28$, $\zeq=3280$ and $\zdec =1090$.  We begin the numerical integration at $a_i = 10^{-10}$; Taylor expansions of Eq.~(\ref{mrPsi}) and its derivative with respect to $a$ around $a=0$ were used to set initial conditions for $\Psi(a_i)$ and $\Psi^\prime(a_i)$ in terms of the primordial $\Psip$.  Figure \ref{fig:Psi} also shows the solutions given by Eqs.~(\ref{mrPsi}) and (\ref{lmPsi}) for comparison.  A key feature of the numerical solution is the value of $\Psi$ at decoupling: $\Psi(\taudec) = 0.937 \Psip$, which is larger than the value $\Psi$ attains during matter-domination [$0.9 \Psip$].

\section{CMB anisotropies from superhorizon potential perturbations}
\label{sec:potential}

Since  $kH_0^{-1}\ll1$ for a superhorizon perturbation, it is desirable to expand $\Psi(\tau, \vec{x})$ in powers of $(\vec{k}\cdot\vec{x})$.  We generalize the expansion of a sine wave perturbation $\Psi = \Psi_{\vec{k}} \sin(\vec{k}\cdot\vec{x}+\varpi)$ by considering a superhorizon potential perturbation of the form
\beqa
\Psi(\tau,\vec{x})&=&\Psi_{\vec{k}}(\tau)\big[\sin \varpi_0+ \cos \varpi_1 (\vec{k}\cdot\vec{x}) - \frac{\sin \varpi_2}{2} (\vec{k}\cdot\vec{x})^2 \nonumber \\
&&\qquad\quad -\left.\frac{\cos \varpi_3}{6}(\vec{k}\cdot\vec{x})^3+ {\cal{O}}(k^4x^4)\right].
\label{PsiExp}
\eeqa
If the potential perturbation is a single sine wave, as would result from a sinusoidal fluctuation in the inflaton field, then all the $\varpi_i$ phases are equal and correspond to the phase of the wave $\varpi$.  We use a more general expression here because it will be useful when considering curvaton perturbations in the next Section.

The expansion in powers of $(\vec{k}\cdot\vec{x})$ of the CMB temperature anisotropy due to the SW effect follows directly from Eqs.~(\ref{SW}) and (\ref{PsiExp}).  The corresponding expansion of the ISW effect is a little more involved.  We start by rewriting the ISW term in Eq.~(\ref{SWandISW}) as
\beq
\left[\frac{\Delta T}{T}(\hat{n})\right]_\mathrm{ISW} = 2\int_{\adec}^{1} \frac{\drm \Psi}{\drm a}\left[a, H_0^{-1}\{\chi_0-\chi(a)\}\hat{n}\right] \,\, \drm a,
\eeq
where 
\beq
\chi(a) \equiv H_0[\tau(a)-\taudec] = \int_{\adec}^a {\frac{\drm a^\prime}{(a^\prime)^2 H(a^\prime)/H_0}},
\eeq
and $\chi_0 \equiv \chi(a=1) = H_0 \xdec$.  We then use Eq.~(\ref{PsiExp}) to expand the integrand in powers of $(\vec{k}\cdot\vec{x})$.   The resulting expression for the ISW effect is
\beqa
\frac{[\Delta T/T]_\mathrm{ISW}}{\Psi_{\vec{k}}(\taudec)}&=&\scri0\sin \varpi_0 
+(\scri0-\scri1) \cos \varpi_1 (\vec{k}\cdot\vxdec)\nonumber\\
&&-(\scri0-2\scri1+\scri2) \frac{\sin \varpi_2}{2} (\vec{k}\cdot\vxdec)^2 \label{Tisw}\\
&&-(\scri0-3\scri1+3\scri2-\scri3)\frac{\cos \varpi_3}{6}(\vec{k}\cdot\vxdec)^3\nonumber,
\eeqa
where $\vxdec \equiv \hat{n}\xdec$ and we have defined
\beq
{\cal I}_n \equiv \frac{2}{(H_0\xdec)^n}\int_{\adec}^{1}\frac{\Psi_{\vec{k}}^\prime(a)}{\Psi_{\vec{k}}(\taudec)}\chi^n(a) \,\, \drm a.
\eeq

Finally, we need to expand the Doppler effect in powers of  $(\vec{k}\cdot\vec{x})$.  Recall from Eq.~(\ref{doppler}) that, to first order in $v$, $[\Delta T/T]_\mathrm{D} = [\vec{v}(\tau_0, \vec{0})-\vec{v}(\taudec, \vxdec)]\cdot \hat{n}$.  Since the time and spatial dependence of $\Psi$ are separable, as shown in Eq.~(\ref{PsiExp}), we can isolate the time dependence of $\vec{v}(\tau,\vec{x})$ by defining a dimensionless quantity
\beqa
{\cal V}(\tau) &\equiv& \frac{-2a^2}{\xdec H_0 \omegaM}\frac{H(a)}{H_0}\left(\frac{y}{4+3y}\right) \nonumber\\
&&\times\left[\frac{\Psi_{\vec{k}}(a)}{\Psi_{\vec{k}}(\taudec)}+\frac{\drm}{\drm \ln a} \frac{\Psi_{\vec{k}}(a)}{\Psi_{\vec{k}}(\taudec)}\right].
\label{calVdef}
\eeqa
It follows from Eq.~(\ref{veqn}) for $\vec{v}(\tau,\vec{x})$ and Eq.~(\ref{PsiExp}) for $\Psi$ that 
\beq
\vec{v}(\tau,\vec{x})=\xdec{\cal V}(\tau) \vec{\nabla} \Psi(\taudec,\vec{x}).
\eeq
Since $\vec{\nabla} \Psi (\taudec, \vec{0}) =  \vec{k} \Psi_{\vec{k}}(\taudec)\cos \varpi_1$, our current velocity only contributes to the $(\vec{k}\cdot\vec{x})$ term of the expansion.  Meanwhile, 
\beqa
 \vec{\nabla} \Psi (\taudec, \vxdec)&=&  \vec{k} \Psi_{\vec{k}}(\taudec) \big[\cos \varpi_1 - \sin \varpi_2 (\vec{k}\cdot\vxdec) \nonumber\\
 &&\qquad\qquad\quad- \left. \frac{\cos \varpi_3}{2}(\vec{k}\cdot\vxdec)^2\right],
 \eeqa
so the velocity of the fluid at the surface of last scattering contributes to all terms in the 
$(\vec{k}\cdot\vec{x})$ expansion.  The temperature anisotropy due to the Doppler effect is therefore
\beqa
\frac{1}{\Psi_{\vec{k}}(\taudec)}\left[\frac{\Delta T}{T}(\hat{n})\right]_\mathrm{D}&=&[{\cal V}(\tau_0)-{\cal V}(\taudec)] \cos \varpi_1 (\vec{k}\cdot\vxdec)\nonumber\\
&&+2{\cal V}(\taudec) \frac{\sin \varpi_2}{2} (\vec{k}\cdot\vxdec)^2\nonumber\\
&&+3{\cal V}(\taudec)\frac{\cos \varpi_3}{6}(\vec{k}\cdot\vxdec)^3.\label{Tdop}
\eeqa

Combining the SW effect, the ISW effect and the Doppler effect gives the total CMB temperature anisotropy produced by a potential perturbation of the form given in Eq.~(\ref{PsiExp}):
\beqa
\frac{\Delta T}{T}(\hat{n}) &=& \Psi_{\vec{k}}(\taudec)\big[\mu(k\xdec)\delta_1\cos\varpi_1\nonumber\\
&&\qquad\qquad- \mu^2(k\xdec)^2\delta_2\frac{\sin \varpi_2}{2} \nonumber\\
&&\qquad\qquad \left.- \mu^3(k\xdec)^3\delta_3\frac{\cos \varpi_3}{6} \right]
\label{delT}
\eeqa
where $\mu \equiv \hat{k}\cdot\hat{n}$. We have discarded the monopolar components of the SW and ISW effects since they only shift the mean CMB temperature and are therefore unobservable.  The $\delta_i$ are
\beqa 
\delta_1 &=& {\cal S}+(\scri0-\scri1)+[{\cal V}(\tau_0)-{\cal V}(\taudec)]  \label{delta1} \\
\delta_2 &=& {\cal S}+(\scri0-2\scri1+\scri2)-2{\cal V}(\taudec)  \\
\delta_3 &=& {\cal S}+(\scri0-3\scri1+3\scri2-\scri3)-3{\cal V}(\taudec)
\eeqa
where, from Eq.~(\ref{SW}), we have defined 
\beqa
{\cal S} \equiv 2-\frac{5}{3}\left[\frac{\frac{9}{10}\Psip}{\Psi(\taudec)}\right]
\eeqa
to be the SW effect's contribution to the anisotropy.  The contribution of the ISW effect to the $\delta_i$ coefficients is contained in the ${\cal I}_n$ terms and follows from Eq.~(\ref{Tisw}).  Finally, the ${\cal V}$ terms are the contribution from the Doppler effect and follow from Eq.~(\ref{Tdop}).

\begin{figure}
\includegraphics[width=8.5cm]{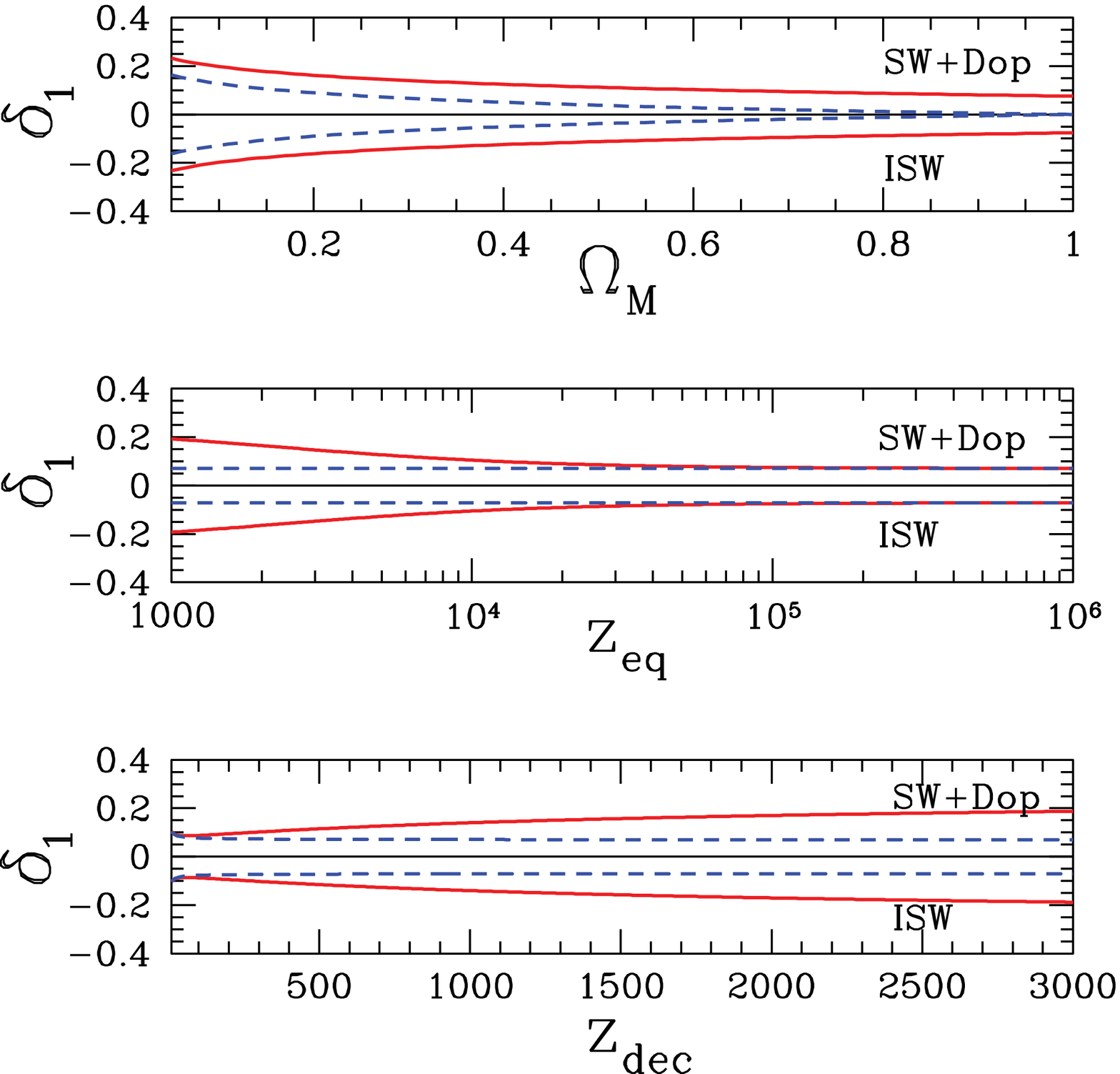}
\caption{The contributions to $\delta_1$, as defined in Eqs.~(\ref{delT}) and (\ref{delta1}), from the Sachs-Wolfe (SW) effect, the Doppler (Dop) effect, and the integrated Sachs-Wolfe (ISW) effect.  The solid curves correspond to a \lcdm universe with radiation (fiducial parameters $\omegaM=0.28$, $\zeq=3280$, and $\zdec=1090$).  The dashed curves correspond to a \lcdm universe with no radiation.  In all cases, the SW+Dop effect is exactly cancelled by the ISW effect so that $\delta_1=0$.  An approximate solution for $\Psi(a)$, accurate to within 0.05\%, was used to generate this figure.}
\label{fig:Dipole}
\end{figure}

In an Einstein-de Sitter universe, the dipole induced by a superhorizon perturbation through the SW effect is exactly cancelled by the Doppler dipole \cite{GZ78, Turner91, BL94}; since there is no ISW effect, we have $\delta_1 = 0$ in this case.   We find that this cancellation extends to flat universes that contain radiation and a cosmological constant in addition to matter, provided that the superhorizon perturbation is adiabatic.  Figure \ref{fig:Dipole} shows the SW+Doppler and ISW terms in $\delta_1$; in the presence of radiation and $\Lambda$, the SW+Doppler term is no longer zero, but it is equal and opposite to the ISW term for all values of $\omegaM$, $\zeq$, and $\zdec$.  Figure~\ref{fig:Dipole} also shows the contributions to $\delta_1$ for a \lcdm universe with no radiation; an analytic derivation of $\delta_1 = 0$ in the absence of radiation is given in Appendix \ref{app:cancel}.  

The cancellation of the ${\cal O}(k\xdec)$ temperature anisotropies also occurs if the universe contains a fluid other than matter or radiation, as shown in Appendix \ref{app:cancel2}, which leads us to suspect that $\delta_1=0$ is a ubiquitous feature of adiabatic perturbations in flat cosmologies.  This conclusion is also supported by going to synchronous gauge, in which galaxies have zero peculiar velocity by construction.  As shown in Ref.~\cite{HS05}, any individual observer in a flat spacetime with a superhorizon adiabatic perturbation can choose Riemann normal coordinates consistent with synchronous gauge such that the metric in the neighborhood of that observer is described by an unperturbed FRW metric plus corrections that are ${\cal O}(k^2 H_0^{-2})$.  Since there are no ${\cal O}(k)$ perturbations and the observer has no extragalactic peculiar velocity in this gauge and therefore sees no Doppler effects due to the superhorizon perturbation, we would not expect the observer to see a temperature anisotropy that is proportional to ${\cal O}(k\xdec)$.  It follows that the ${\cal O}(k\xdec)$ temperature anisotropies in any other gauge must sum to zero.  

Since $\delta_1=0$, the leading order anisotropy will be the $\delta_2$ term, unless $\sin \varpi_2=0$, in which case the $\delta_3$ term will be the leading order contribution.  For a flat \lcdm Universe with $\omegaM=0.28$, $\zeq=3280$, and $\zdec=1090$, $\delta_2 = 0.33$ and $\delta_3=0.35$.   

As a brief aside, let us consider the $v_\mathrm{net}^2$ term in Eq.~(\ref{doppler}) for the Doppler effect.    Like the $\delta_2$ term in Eq.~(\ref{delT}), this term is proportional to $(\vec{k}\cdot\vxdec)^2$.  
Since $v_\mathrm{net}$ could be near unity if the gradient of the superhorizon perturbation is large, we may be concerned that this term will produce a quadrupolar anisotropy that is comparable to, or even larger than, the quadrupole given by the $\delta_2$ term in Eq.~(\ref{delT}).  Fortunately, this concern is unfounded because the Doppler shift of the intrinsic dipole also produces a quadrupole: if there is an intrinsic temperature pattern $T_\mathrm{int}(\hat{n})$ in the CMB, then an observer moving with velocity $\vec{v}_\mathrm{net}$ with respect to the CMB will measure a temperature pattern $T_\mathrm{int}(\hat{n})/[\gamma(1-\vec{v}_\mathrm{net}\cdot\hat{n})]$.  Taylor expanding in $v_\mathrm{net}$ and dropping terms that are isotropic gives the observed temperature fluctuation if the CMB has an intrinsic anisotropy:
\beqa
\frac{T_\mathrm{obs}}{T} (\hat{n}) &=&\left[1 + \frac{\Delta T_\mathrm{int}(\hat{n})}{T}\right] \left[1+\vec{v}_\mathrm{net}\cdot \hat{n}+v_\mathrm{net}^2(\hat{v}_\mathrm{net}\cdot\hat{n})^2 \right], \nonumber\\
&=& 1+(\vec{k}\cdot\vxdec)\delta_1 \Psi_{\vec{k}}(\taudec)\cos\varpi_1+  \ldots +\nonumber\\
&&(\vec{k}\cdot\vxdec)^2\delta_1[{\cal V}(\tau_0)-{\cal V}(\taudec)]\Psi^2_{\vec{k}}(\taudec)\cos^2\varpi_1,\nonumber
\eeqa
where the second line follows from $\Delta T_\mathrm{int}(\hat{n})/T \simeq (\vec{k}\cdot\vxdec) \Psi_{\vec{k}}(\taudec)\cos\varpi_1[{\cal S}+\scri0-\scri1]$ and the ellipses contain the linear quadrupolar and octupolar anisotropies given in Eq.~({\ref{delT}}).  Given that $\delta_1=0$, we see that the Doppler quadrupole is exactly cancelled by the Doppler-shifted intrinsic dipole.  Of course, this Doppler quadrupole is nonlinear in $\Psi$, so it is questionable to analyze it using linear theory.  Nevertheless, this argument shows that this term is no more alarming than any other higher-order term in $\Psi$.

To compare the CMB anisotropy given by Eq.~(\ref{delT}) to observations, we must decompose this anisotropy into multipole moments:
\beq
\frac{\Delta T}{T}(\hat{n}) = \sum_{\ell,m} a_{\ell m} Y_{\ell m}(\hat{n}).
\eeq
Given the addition theorem of spherical harmonics,
\beq
P_\ell(\mu) = \frac{4\pi}{2\ell+1}\sum_{m=-\ell}^{\ell} Y_{\ell m}^*(\hat{k})Y_{\ell m}(\hat{n}),
\eeq
the values of $a_{1m}$, $a_{2m}$ and $a_{3m}$ are easily obtained from Eq.~(\ref{delT}).  It is also clear that each $a_{\ell m}$ is proportional to $Y_{\ell m}^*(\hat{k})$.  Consequently, if $\hat{k}$ is chosen to lie on the $z$ axis, then the only nonzero moments are $a_{10}$, $a_{20}$ and $a_{30}$.  In this case, with $\delta_1=0$,
\beqa
a_{10} &=& -\sqrt{\frac{4\pi}{3}} (k\xdec)^3 \delta_3 \frac{\cos \varpi_3}{10} \Psi_{\vec{k}}(\taudec),\\
a_{20} &=& -\sqrt{\frac{4\pi}{5}} (k\xdec)^2 \delta_2 \frac{\sin \varpi_2}{3} \Psi_{\vec{k}}(\taudec), \label{a20}\\
a_{30} &=& -\sqrt{\frac{4\pi}{7}} (k\xdec)^3 \delta_3 \frac{\cos \varpi_3}{15}  \Psi_{\vec{k}}(\taudec) \label{a30}.
\eeqa

Thus we see that even though $\delta_1=0$, a superhorizon potential perturbation still induces a dipolar anisotropy in the CMB.  However, this anisotropy is suppressed by a factor of $(k\xdec)^3$.  Moreover, it is comparable in magnitude to the induced octupolar anisotropy.  Since measurements of $|a_{10}|$ are contaminated by our peculiar velocity, the upper bound on $|a_{10}|$ ($|a_{10}|\lsim 10^{-3}$) is much higher than the upper bound on $|a_{30}|$.  Therefore, the most restrictive constraints on $\Psi_{\vec{k}}(\taudec)$ come from Eqs.~(\ref{a20}) and (\ref{a30}):
\beqa
     (k \xdec)^2 \left|\Psi_{k}(\taudec)\sin \varpi_2  \right|
     &\lesssim& 5.8 \, Q \label{quad}\\
     (k \xdec)^3 \left|\Psi_{k}(\taudec)\cos \varpi_3 \right|
     &\lesssim& 32 \,  {\cal O} \label{oct}
\eeqa
where $Q$ and ${\cal O}$ are upper bounds on $|a_{20}|$ and $|a_{30}|$, respectively, in a coordinate system aligned with the superhorizon perturbation $(\hat{k} = \hat{z})$.  

Since other primordial perturbations, including smaller-scale modes, may also contribute to the measured values of $|a_{20}|$ and $|a_{30}|$ in a way that suppresses the perturbation from the single superhorizon mode we have been considering, the $|a_{20}|$ and $|a_{30}|$ values from the superhorizon mode may be as large as the largest values of $|a_{20}|$ and $|a_{30}|$ that are consistent with the measured variance in these moments.   We take $Q = 3\sqrt{C_2} \lesssim 1.8\times10^{-5}$ and ${\cal O} = 3\sqrt{C_3} \lesssim 2.7 \times 10^{-5}$, three times the measured rms values of the quadrupole and octupole \cite{Efstathiou04}, as $3\sigma$ upper limits.

When a superhorizon potential perturbation is invoked to generate a power asymmetry in the CMB \cite{EKC08}, it is the variation of $\Psi$ across the surface of last scattering, $\Delta \Psi (\taudec) \simeq |\Psi_{k}(\taudec)  (k \xdec) \cos \varpi_1|$, that is the relevant quantity.  For a given $\Delta \Psi(\taudec)$, the induced CMB quadrupole and octupole can be made arbitrarily small by decreasing $(k\xdec)$.  However, if the superhorizon perturbation is a single mode of the form $\Psi = \Psi_{\vec{k}} \sin(\vec{k}\cdot\vec{x}+\varpi)$, then demanding that $\Psi\lsim1$ everywhere, even outside our Hubble volume, leads to an additional constraint: $\Psi_{\vec{k}}(\taudec) \lsim 1$ implies that $(k\xdec) \gsim \Delta \Psi/\cos \varpi$.  Moreover, since $\varpi_2=\varpi_3=\varpi$ in this case, the bounds given by Eqs.~(\ref{quad}) and (\ref{oct}) imply that
\beqa
\Delta\Psi (\taudec) (k \xdec) \left|\tan \varpi\right| &\lesssim& 5.8 \, Q \label{quadDel} \\
  \Delta\Psi (\taudec) (k \xdec)^2 &\lesssim& 32 \,  {\cal O}  \label{octDel}.
\eeqa

Since the octupole constraint on $\Delta\Psi (\taudec)$ is independent of $\varpi$ while the quadrupole constraint vanishes if $\varpi=0$, the maximum allowed value for $\Delta \Psi (\taudec)$ is obtained when $\varpi =0$.  In this case, we have 
\beq
\Delta \Psi (\taudec) \lsim \mbox{Min}\left[(k\xdec),\frac{32 \,  {\cal O}}{(k\xdec)^2}\right],
\eeq
where the first bound follows from $\Psi_{\vec{k}}(\taudec) \lsim 1$ and the second follows from Eq.~(\ref{octDel}).  Consequently, the maximum value for $\Delta \Psi (\taudec)$ is obtained when $\Delta \Psi (\taudec) \simeq [32 \,  {\cal O}]^{1/3} \simeq 0.095$.  In this case, $\Delta \Psi (\taudec) \simeq (k\xdec)$, which implies that the wavelength of the superhorizon mode is 65 times larger than the particle horizon.  This is much smaller than the lower bound on the wavelength of order-unity perturbations found in Ref. \cite{CDF03} because we have placed ourselves at the node ($\varpi =0$) of a single-mode perturbation, thus eliminating the induced quadrupolar anisotropy that would have otherwise provided a much more stringent constraint.  Nevertheless, the octupole constraint alone is sufficient to rule out an inflaton perturbation large enough to generate the observed power asymmetry in the CMB \cite{EKC08}. 

\section{Application to Curvaton Perturbations}
\label{sec:curvaton}
The constraints given by Eqs. (\ref{quad}) and (\ref{oct}) may also be applied to superhorizon perturbations that are not describable by a single sine wave fluctuation in $\Psi$.  In this section, we will show how these constraints limit the amplitude of a superhorizon fluctuation in a curvaton field.   Superhorizon curvaton fluctuations may be a generic feature of the curvaton model of inflation \cite{LM06}, and a superhorizon curvaton fluctuation is capable of generating the observed power asymmetry in the CMB \cite{EKC08}.

In the curvaton model \cite{Mollerach90, LM97, LW02, MT01}, there are two fields present during inflation: the inflaton dominates the energy density of the Universe and drives inflation while the curvaton ($\sigma$) generates some or all of the primordial perturbations.  The curvaton potential is assumed to be $V(\sigma) = (1/2)m_\sigma^2 \sigma^2$, with $m_\sigma \ll H_I$, where $H_I $ is the value of the Hubble parameter during inflation.  Consequently, the curvaton is effectively massless during inflation and remains frozen at its initial value.  After inflation ends and $m_\sigma \simeq H$, the curvaton will oscillate about its minimum, behaving like a cold gas of $\sigma$ particles.  The curvaton is then assumed to decay into radiation prior to neutrino decoupling, generating a gauge-invariant curvature perturbation \cite{BST83, BTW06} $\zeta \simeq (R/3) \delta \rho_\sigma/\rho_\sigma$, where $R\equiv(\rho_\sigma/\rho_\mathrm{tot})$ is the fraction of the total energy density in the curvaton field just prior to its decay.  

During radiation domination, a curvature perturbation $\zeta$ corresponds to a potential perturbation $\Psi = -(2/3)\zeta$.  We assume that the curvaton decay occurred early enough that $\Psip \simeq - (2R/9)\delta \rho_\sigma/\rho_\sigma$.  In this case, a perturbation in the curvaton field induces a potential perturbation at decoupling given by 
\beq
\Psi(\taudec) \simeq - \frac{R}{5}\left[\frac{\Psi(\taudec)}{\frac{9}{10}\Psip}\right]\left[2
     \left(\frac{\delta\sigma}{\bar\sigma}\right) +
     \left(\frac{\delta\sigma}{\bar\sigma}\right)^2
     \right],
\label{sigmaPsi}
\eeq  
where $\bar\sigma$ is the spatially homogeneous background value of the curvaton field and $\sigma(\vec{x}) = \bar\sigma + \delta\sigma(\vec{x})$.  We consider a superhorizon sinusoidal perturbation to the curvaton field $\delta\sigma = \sigma_{\vec{k}} \sin(\vec{k}\cdot\vec{x}+\varpi_\sigma)$.  Since adding $\pi$ to $\varpi_\sigma$ changes the sign of $\delta\sigma$, we may assume that $\sigma_{\vec{k}} /\bar\sigma > 0$ without loss of generality.  The potential perturbation induced by $\delta\sigma$ may be expanded in powers of $(\vec{k}\cdot\vec{x})$ and the resulting expression is Eq.~(\ref{PsiExp}) with
\beqa
\sin \varpi_0 &=& \sin \varpi_\sigma + \left(\frac{\sigma_{\vec{k}}}{2\bar\sigma}\right) \sin^2 \varpi_\sigma, \\
\cos \varpi_1 &=& \cos \varpi_\sigma + \left(\frac{\sigma_{\vec{k}}}{2\bar\sigma}\right) \sin 2\varpi_\sigma, \\
\sin \varpi_2 &=& \sin \varpi_\sigma - \left(\frac{\sigma_{\vec{k}}}{\bar\sigma}\right) \cos 2\varpi_\sigma, \\
\cos \varpi_3 &=& \cos \varpi_\sigma+2\left(\frac{\sigma_{\vec{k}}}{\bar\sigma}\right)\sin 2\varpi_\sigma.
\eeqa
Since $a_{20} \propto \sin \varpi_2$, we can see that it will not be possible to choose a phase $\varpi_\sigma$ such that the CMB quadrupole vanishes for all values of $\sigma_{\vec{k}}$.  For specific values of $\sigma_{\vec{k}}$, however, there will be values of $\varpi_\sigma$ for which $\sin \varpi_2 = 0$.

The constraint from the CMB quadrupole [Eq.~(\ref{quad})] implies
\beqa
&&R\left|\frac{\Delta\sigma}{\bar\sigma}\right|\left|\tan \varpi_\sigma - \frac{\Delta\sigma}{\bar\sigma(k\xdec)} \frac{\cos 2\varpi_\sigma}{\cos^2 \varpi_\sigma}\right| \nonumber \\
&&\qquad\qquad\qquad\qquad \lsim \frac{5}{2}\left(\frac{5.8 \, Q}{k\xdec}\right) \left[\frac{\frac{9}{10}\Psip}{\Psi(\taudec)}\right],
\label{cuvatonQgen}
\eeqa
where $\Delta\sigma = \sigma_{\vec{k}} (k\xdec)\cos \varpi_\sigma$ is the variation in the curvaton field across the surface of last scattering.   If $\varpi_\sigma=0$, then the CMB quadrupole anisotropy is sourced exclusively by the term in Eq.~(\ref{sigmaPsi}) that is proportional to $(\delta\sigma)^2$, and the resulting constraint is independent of $k\xdec$.  In this case, Eq.~(\ref{cuvatonQgen}) reduces to
\beq
R \left(\frac{\Delta\sigma}{\bar\sigma}\right)^2 \lsim \frac{5}{2} (5.6 \, Q) \mbox{ for } \varpi_\sigma =0,
\label{zeroQ}
\eeq
where we have used $\Psi(\taudec) = 0.937 \Psip$, as derived in Section \ref{sec:basics}.  

\begin{figure}
\includegraphics[width=8.5cm]{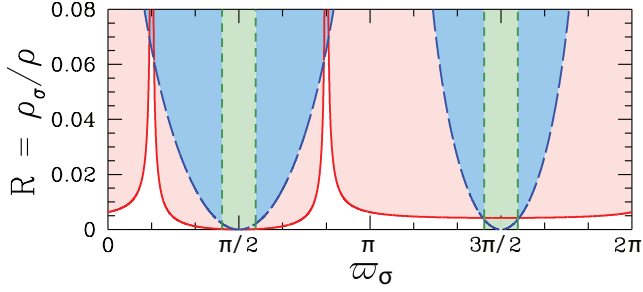}
\caption{The upper bounds on $R$ (the fraction of the energy density in the curvaton field just prior to its decay) for a superhorizon curvaton fluctuation with $\delta\sigma = \bar \sigma \sin(\vec{k}\cdot\vec{x}+\varpi_\sigma)$.  In this plot, $|\Delta \sigma/\bar\sigma|=(k\xdec)|\cos \varpi_\sigma| =0.2$ and the shaded regions are excluded.  The solid curve is the bound placed by the CMB quadrupole, and it vanishes for $\varpi_\sigma = \pi/6$ and $\varpi_\sigma = 5\pi/6$.  The long-dashed curve is the bound from the CMB octupole.  The shaded areas enclosed by the short-dashed lines around $\varpi_\sigma =\pi/2$ and $\varpi_\sigma =3\pi/2$ are not allowed because $(k\xdec)>1$ in these regions.}
\label{fig:phases}
\end{figure}

If $\varpi_\sigma$ is nonzero, then the quadrupole constraint given by Eq.~(\ref{cuvatonQgen}) will depend on $k\xdec$.  For instance, if $\varpi_\sigma = \pi/4$, Eq.~(\ref{cuvatonQgen}) reduces to
\beq
R\left| \frac{\Delta\sigma}{\bar\sigma} \right| \lsim \frac{5}{2}\left(\frac{5.6 \, Q}{k\xdec}\right) \mbox{ for } \varpi_\sigma = \frac{\pi}{4}.
\eeq
This upper bound may be made arbitrarily large by decreasing $k\xdec$.  However, the condition that $\delta \sigma \lsim \bar \sigma$ leads to a lower bound on $k\xdec$:  $k\xdec \gsim \Delta \sigma/(\bar\sigma \cos \varpi_\sigma)$.  We now set $k\xdec = \Delta \sigma/(\bar\sigma \cos \varpi_\sigma)$ (or equivalently $\sigma_{\vec{k}} = \bar \sigma$) and consider how the quadrupole constraint depends on the phase of the curvaton wave.  Figure~\ref{fig:phases} shows the quadrupole bound on $R$ as a function of $\varpi_\sigma$ for $|\Delta \sigma/\bar\sigma|=0.2$.  As mentioned earlier, there are some values of $\varpi_\sigma$ for which there is no induced CMB quadrupole; in Fig.~\ref{fig:phases} we see that setting $\sigma_{\vec{k}} = \bar \sigma$ implies that the quadrupole constraint is lifted if $\varpi_\sigma = \pi/6$ or $5\pi/6$.  We also note that setting $\sigma_{\vec{k}} = \bar \sigma$ implies an additional constraint on $\Delta \sigma/\bar\sigma$: the curvaton perturbation is superhorizon only if $|\Delta \sigma/\bar\sigma| < |\cos \varpi_\sigma|$.  

The CMB octupole constraint implied by Eq.~(\ref{oct}) for a curvaton perturbation is
\beqa
&&R\left|\frac{\Delta\sigma}{\bar\sigma}\right|\left|1+2\frac{\Delta\sigma}{\bar\sigma(k\xdec)}\frac{\sin 2\varpi_\sigma}{\cos^2\varpi_\sigma}\right| \nonumber \\
&&\qquad\qquad\qquad\lsim \frac{5}{2}\left[\frac{32 \, {\cal O}}{(k\xdec)^2}\right] \left[\frac{\frac{9}{10}\Psip}{\Psi(\taudec)}\right],
\label{cuvatonOgen}
\eeqa
and the resulting upper bound on $R$ for $\sigma_{\vec{k}} = \bar \sigma$ and $|\Delta \sigma/\bar\sigma|=0.2$ is shown in Fig.~\ref{fig:phases}.  Since the CMB octupole generated by a superhorizon perturbation is suppressed by a factor of $k\xdec$ relative to the CMB quadrupole, the octupole constraint is weaker than quadrupole constraint for most values of $\varpi_\sigma$.   The only exceptions are $\varpi_\sigma = \pi/6$ or $5\pi/6$ since the quadrupole constraint is lifted for these phases if $\sigma_{\vec{k}} = \bar \sigma$.  In this case, Eq.~(\ref{cuvatonOgen}) implies
\beq
R\left|\frac{\Delta\sigma}{\bar\sigma}\right|^3 \lsim \frac{5}{8}(31 \, {\cal O}) \mbox{ for } \varpi_\sigma = \left\{\frac{\pi}{6} \mbox{ or  } \frac{5\pi}{6} \right\} \mbox { and } \sigma_{\vec{k}} = \bar \sigma,
\eeq
and we see in Fig.~\ref{fig:phases} that this bound on $R$ is far less restrictive than the quadrupole bound for other values of $\varpi_\sigma$.

\begin{figure}
\includegraphics[width=8.5cm]{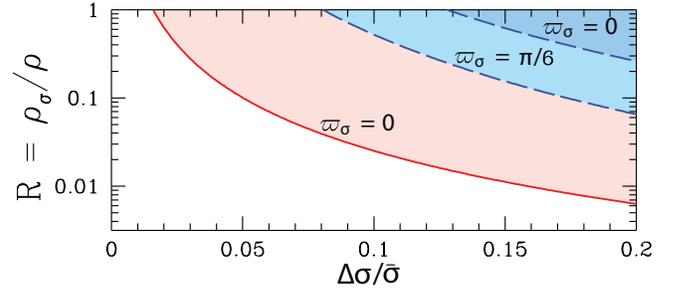}
\caption{The $R$-$\Delta\sigma/\bar\sigma$ parameter space for a superhorizon curvaton fluctuation with $\delta\sigma = \bar \sigma \sin(\vec{k}\cdot\vec{x}+\varpi_\sigma)$; the shaded regions are excluded.  $\Delta \sigma/\bar\sigma$ is the variation of the curvaton field across the surface of last scattering divided by the background value of the cuvaton and $R$ is the same as in Fig.~\ref{fig:phases}.  The solid curve is the bound from the CMB quadrupole, and the long-dashed curves are bounds from the CMB octupole.  The bounds are plotted for two values of $\varpi_\sigma$; there is no quadrupole bound if $\varpi_\sigma=\pi/6$.}
\label{fig:bounds}
\end{figure}

The CMB power asymmetry induced by a superhorizon curvaton fluctuation is proportional to $(\Delta\sigma/\bar\sigma)$.  Consequently, the quadrupole and octupole constraints establish upper bounds on $R$ for a given power asymmetry, as shown in Fig.~\ref{fig:bounds}.  This upper bound increases as the required power asymmetry decreases.  There is also a lower bound on $R$ that arises from limits to non-Gaussianity in the CMB, but there is a small range of $R$ values for which it is possible to generate the observed CMB power asymmetry with a superhorizon curvaton fluctuation even if the induced CMB quadrupole does not vanish \cite{EKC08}.

\section{Summary and Conclusions}
\label{sec:conclusions}
Superhorizon perturbations generate large-scale anisotropies in the CMB through the Grishchuk-Zel'dovich effect \cite{GZ78}.   In this paper, we have derived the constraints to single-mode adiabatic superhorizon perturbations that arise from measurements of the CMB quadrupole and octupole.  These constraints differ from those previously derived for an isotropic distribution of superhorizon perturbations \cite{GZ78, Turner91, KTF94,  GLLW95, CDF03} because the CMB anisotropies generated by a single-mode perturbation depend on the perturbation's phase.

We started by considering a sinusoidal superhorizon gravitational potential perturbation with wavenumber $k\ll H_0$.  Since the leading-order term in the potential perturbation is proportional to $(\vec{k}\cdot\vec{x})$, it would be expected to generate a dipolar anisotropy of comparable amplitude in the CMB through the Sachs-Wolfe effect.  However, the superhorizon perturbation also gives us a velocity with respect to the CMB, and the resulting Doppler dipole exactly cancels the leading-order intrinsic anisotropy generated by the SW and ISW effects, provided that the perturbation is adiabatic.  This cancellation was known to occur in an Einstein-de Sitter universe \cite{GZ78, Turner91, BL94}, but we found that it also applies to flat \lcdm universes with and without radiation, as well as in more exotic flat cosmological models.

Due to this cancellation of the CMB dipole, the leading-order constraints on adiabatic superhorizon fluctuations arise from measurements of the CMB quadrupole and octupole.  If the potential perturbation is sinusoidal, as would be created by a sinusoidal fluctuation in the inflaton, then putting ourselves at the node of the sine wave maximizes the difference in potential across the Universe while also eliminating the induced quadrupole anisotropy in the CMB.  In this case, the CMB octupole provides the strongest constraint on the amplitude of the superhorizon perturbation: $\Delta \Psi \lsim 0.095$, where $\Delta \Psi$ is the variation of the potential $\Psi$ across the surface of last scattering.

A fluctuation in a field that contains only a small fraction of the energy density of the Universe generates a smaller potential perturbation and, consequently, smaller CMB anisotropies.  We consider a multi-field model of inflation in which a subdominant curvaton field generates primordial perturbations \cite{Mollerach90, LM97, LW02, MT01}.   For a given superhorizon fluctuation in the curvaton field, the measured values of the CMB quadrupole and octupole place upper bounds on the fraction of the total energy density contained in the curvaton field prior to its decay.  Since a sinusoidal perturbation in the curvaton field generates a potential perturbation that is not sinusoidal, there is no value for the phase of the curvaton fluctuation that eliminates the induced CMB quadrupole for any superhorizon curvaton fluctuation.  However, once the amplitude of the curvaton perturbation is specified, it is possible to choose a phase for which the induced CMB quadrupole vanishes.  In this case, measurements of the CMB octupole still place an upper bound on the curvaton energy density, but this bound is significantly weaker than the bound from the CMB quadrupole that applies to curvaton fluctuations with different phases.

Superhorizon perturbations have generated interest recently because they are a simple way to introduce a preferred direction in the Universe and may generate the deviations from statistical isotropy that have been observed in the CMB.  In particular, in Ref. \cite{EKC08}, we showed that a superhorizon perturbation to an inflationary field can generate the hemispherical power asymmetry found in the WMAP data \cite{Eriksen04, HBG04, Eriksen07}.   In this paper, we have demonstrated how the CMB constrains such superhorizon perturbations: the octupole constraint on $\Delta \Psi$ is sufficient to rule out an inflaton perturbation as the source of the observed power asymmetry, but it is possible to generate the observed power asymmetry with a superhorizon curvaton perturbation.  These constraints may also be applied to other scenarios that invoke superhorizon perturbations.  For instance, order-unity superhorizon fluctuations in the mean value of the curvaton may be a generic feature of the curvaton model \cite{LM06}.

\begin{acknowledgments}
We thank Chris Hirata and Mike Kesden for useful discussions.  This work was supported by DoE DE-FG03-92-ER40701 and the Gordon and Betty Moore Foundation.
\end{acknowledgments}

\appendix
\section{Dipole Cancellation in a \lcdm Universe}
\label{app:cancel}
In this Appendix, we show how the intrinsic CMB dipole induced by a superhorizon adiabatic perturbation is exactly cancelled by the dipole arising from the Doppler effect in a \lcdm Universe with negligible radiation.  Specifically, we will assume that decoupling occurred long after matter-radiation equality so that the evolution of the gravitational potential is given by Eq. (\ref{lmPsi}). 

First, we derive an alternate expression for the dipole produced by the ISW effect.  For a superhorizon mode with $\Psi(\tau,\vec{x}) \simeq \Psi_{\vec{k}}(\tau)[\vec{k}\cdot\vec{x}]$, the dipolar component of the ISW anisotropy is given by 
\beqa
\left[\frac{\Delta T}{T}(\hat{n})\right]_\mathrm{ISW} &=& 2\int_{\taudec}^{\tau_0} \frac{\drm \Psi_{\vec{k}}}{\drm \tau}\vec{k}\cdot\hat{n}\left[\tau_0-\tau\right] \drm \tau, \\
&=& \left(\vec{k}\cdot\hat{n}\xdec\right) \Psi_{\vec{k}}(\taudec) \nonumber\\
&&\times\left[-2 + \frac{2}{\xdec}\int_{\taudec}^{\tau_0} \frac{\Psi_{\vec{k}}(\tau)}{\Psi_{\vec{k}}(\taudec)} \drm \tau\right], \nonumber
\eeqa
where we integrated by parts to obtain the second line.  In subsequent expressions, we will omit the factor of $(\vec{k}\cdot\hat{n}\xdec) \Psi_{\vec{k}}(\taudec) \simeq \Psi(\taudec,\hat{n}\xdec)$.

Transforming the integral over $\tau$ into an integral over $a$ and using Eq. (\ref{lmPsi}) for $\Psi_{\vec{k}}(a)$ gives
\beq
\left[\frac{\Delta T}{T}\right]_\mathrm{ISW} = -2 + \left[\frac{\frac{9}{10}\Psikp}{\Psi_{\vec{k}}(\taudec)}\right]\frac{5\omegaM}{H_0\xdec}\int_{\adec}^{1} \frac{G(a)}{a^3} \drm a
\label{basicisw}
\eeq
where $G(a)$ is defined by
\beq
G(a) \equiv H_0^3 \int_0^a \frac{\drm a^\prime}{[a^\prime H(a^\prime)]^3} = \int_0^a{\frac{\drm a^\prime}{\left[a^\prime\sqrt{\omegaM(a^\prime)^{-3}+\omegaL} \right]^3}}.
\label{Gdef2}
\eeq
Integrating Eq.~(\ref{basicisw}) by parts yields
\beqa
\left[\frac{\Delta T}{T}\right]_\mathrm{ISW} = -2 &+&\left[\frac{\frac{9}{10}\Psikp}{\Psi_{\vec{k}}(\taudec)}\right]\frac{5\omegaM}{H_0\xdec} \label{delTisw} \\
&\times&\left[\frac{G(\adec)}{2\adec^{2}} - \frac{G_0}{2}+ \half \int_{\adec}^{1} \frac{H_0^3\,\,\drm a}{a^5 H^3(a)}\right],\nonumber
\eeqa
where $G_0 \equiv G(a=1)$.  Since decoupling occurs long before matter-$\Lambda$ equality, $\Psi_{\vec{k}}(\taudec)\simeq [9/10] \Psikp$.  However, we wish to show that the dipole cancellation applies more generally, and so we do not assume that the universe is matter-dominated at decoupling.

The next step is crucial for the upcoming cancellation, and it relies on a special feature of $G(a)$.   Focusing on the last term of Eq.~(\ref{delTisw}), we see that
\beqa
\int_{\adec}^{1} \frac{H_0^3\,\,\drm a}{a^5 H^3(a)} &=& \int_{\adec}^{1} \frac{\drm a}{a^5\left[\omegaM a^{-3}+\omegaL\right]^{3/2} }, \nonumber \\
&=& \frac{2}{3\omegaM} \int_{\adec}^{1} \frac{1}{a} \frac{\drm}{\drm a} \left[ \frac{1}{\sqrt{\omegaM a^{-3} +\omegaL}}\right]\drm a, \nonumber \\
&=& \frac{2}{3\omegaM}\left[1-\frac{1}{\adec}\frac{H_0}{H(\adec)}+H_0 \xdec\right]. \nonumber
 \eeqa
Using this result to eliminate the integral in Eq.~(\ref{delTisw}) gives
\beqa
\left[\frac{\Delta T}{T}\right]_\mathrm{ISW} &=&-2+\frac{5}{3}\left[\frac{\frac{9}{10}\Psikp}{\Psi_{\vec{k}}(\taudec)}\right]+\frac{1}{H_0\xdec}\left[\frac{\frac{9}{10}\Psikp}{\Psi_{\vec{k}}(\taudec)}\right] \nonumber\\
&&\times\left[\frac{H_0 D(\adec)}{H(\adec)\adec^2}-D_0+\frac{5}{3}-\frac{5H_0}{3\adec H(\adec)}\right],
\nonumber\\
\label{finalisw}
\eeqa
where $D(a) \equiv (5/2)\omegaM [H(a)/H_0] G(a)$ so that $\Psi(a) = [9/10]\Psikp D(a)/a$ in Eq.~(\ref{lmPsi}), and $D_0 \equiv D(a=1)$.  The first two terms in this expression are exactly cancelled by the anisotropy produced by the SW effect, which was given in Eq.~(\ref{SW}).  We will now show that the last term in the ISW anisotropy will be cancelled by the Doppler effect.

The Doppler dipole is 
\beq
\left[\frac{\Delta T}{T}(\hat{n})\right]_\mathrm{Doppler} =\left(\vec{k}\cdot\hat{n}\xdec\right) \Psi_{k}(\taudec)\left[{\cal V}(\tau_0)-{\cal V}(\taudec)\right].
\eeq
Long after matter-radiation equality ($y\rightarrow \infty$), with $\Psi(a) = [9/10]\Psikp D(a)/a$, Eq.~(\ref{calVdef}) for ${\cal V}(\tau)$ becomes
\beq
{\cal V}(\tau) = -\left[\frac{\frac{9}{10}\Psikp}{\Psi_{\vec{k}}(\taudec)}\right]\frac{2a(\tau)^2}{3 \omegaM H_0\xdec} \left[\frac{H(\tau)}{H_0}\right] \left.\frac{\drm D}{\drm a}\right|_{a(\tau)}.
\label{genveqn}
\eeq
From
\beq
\frac{\drm D}{\drm a} = -\frac{3\omegaM}{2a^3}\left[\frac{H_0^2}{H^2(a)}\right]\left[\frac{D(a)}{a}-\frac{5}{3}\right]
\eeq
it follows that
\beqa
{\cal V}(\tau_0) &-& {\cal V}(\taudec)= \frac{1}{H_0\xdec}\left[\frac{\frac{9}{10}\Psikp}{\Psi_{\vec{k}}(\taudec)}\right] \nonumber\\
&\times&\left[D_0-\frac{5}{3}-\frac{H_0 D(\adec)}{H(\adec)\adec^2}+\frac{5H_0}{3\adec H(\adec)}\right].
\nonumber
\label{vfulleqn}
\eeqa
As promised, this contribution to the anisotropy cancels the last term in Eq.~(\ref{finalisw}).  Thus we see that the CMB temperature anisotropy terms proportional to $\hat{k} \cdot \hat{n}\xdec$ arising from the SW, ISW and Doppler effects sum to zero in a \lcdm universe.  Consequently, the CMB temperature dipole is comparable in magnitude to the temperature octupole since both are primarily sourced by the component of the temperature anisotropy that is proportional to $(\hat{k} \cdot \hat{n}\xdec)^3$.  

The analytical calculation presented in this Appendix does not apply if there is a significant amount of radiation at decoupling.  The presence of radiation would change the evolution of $\Psi$, preventing us from relating the integral in Eq.~(\ref{basicisw}) to $\xdec$.  However a numerical calculation confirms that the same dipole cancellation occurs when radiation is present, as discussed in Section \ref{sec:potential} and illustrated in Fig.~\ref{fig:Dipole}.

\section{Dipole Cancellation in a Universe with an Exotic Fluid}
\label{app:cancel2}
To demonstrate that the CMB dipole cancellation is not a special feature of universes containing only matter and radiation, we calculate $\delta_1$ for adiabatic superhorizon perturbations in a flat universe that contains an $X$ fluid with constant equation of state $w = p_X/\rho_X$ and a cosmological constant.  In this two-component universe, the Hubble parameter is 
\beq
H^2(a) = H_0^2\left[\frac{\omegaX}{a^{3(1+w)}}+\omegaL\right]
\label{HeqnX}
\eeq
where $\omegaX$ is the present day ratio of $\rho_X$ and the critical density and $\omegaL=1-\omegaX$.  Of course, the existence of the CMB in this universe implies that there is some radiation that we have not included in Eq.~(\ref{HeqnX}).  To justify this omission, we will only consider values of $w$ that are greater than 1/3 so that the $X$-fluid energy density is always greater than the radiation density, even in the very early universe.

We now consider an adiabatic superhorizon perturbation in this universe.  The perturbed fluid, which  contained matter and radiation in Section \ref{sec:basics}, is now dominated by a single component: the $X$ fluid.  The overdensity of the $X$ fluid in its rest frame is therefore related to its peculiar motion through Eqs.~(\ref{vdelta1}) and (\ref{vdelta2}), with $w$ being the equation of state parameter of the $X$ fluid.  The potential perturbation in a flat universe is directly related to the density perturbation \cite{HS95}:
\beq
\Psi_{\vec{k}} = \frac{-4\pi G}{k^2} a^2 \rho \Delta_{\vec{k}},
\label{PsiDeltaX}
\eeq
where $\rho$ is the density of the perturbed fluid.  Setting $\rho$ equal to the sum of the matter and radiation densities yields Eq.~(\ref{PsiDelta}), but in this case the only component of the perturbed fluid is the $X$ fluid, so $\rho = \rho_X = \omegaX [3H_0^2/(8\pi G)] a^{-3(1+w)}$.   We can use Eqs. (\ref{vdelta1}), (\ref{vdelta2}), and (\ref{PsiDeltaX}) to derive an equation for containing only $\Psi_{\vec{k}}(a)$, just as we did in Section \ref{sec:basics}.  The resulting equation is
\beqa
&0&=\Psi^{\prime\prime}_{\vec{k}}(a) + \frac{1}{a}\left[5+3w+\frac{\drm \ln H}{\drm \ln a}\right]\Psi^\prime_{\vec{k}}(a) \label{PsieqnX} \\
&+& \frac{1}{a^2}\left[3(1+w)+\frac{\drm \ln H}{\drm \ln a} - \frac{3(1+w)}{2}\left\{ \frac{\omegaX}{a^{3(1+w)}} \right\}\frac{H_0^2}{H^2}\right]\Psi_{\vec{k}}.\nonumber
\eeqa
We may also use Eq.~(\ref{PsiDeltaX}) to eliminate $\Delta_{\vec{k}}$ from Eq.~(\ref{vdelta1}), which yields an equation for $\vec{v}$:
\beq
\vec{v}(\tau,\vec{x}) = -\frac{2a^{3w+2}}{3(1+w)}\frac{1}{H_0 \omegaX}\frac{H(a)}{H_0}\left[\gradPsi+\frac{\drm}{\drm \ln a} \gradPsi\right].
\label{veqnX}
\eeq

In the very early universe, the $X$ fluid's energy density is much greater than the vacuum energy, and we may neglect $\omegaL$ in Eq.~(\ref{HeqnX}) for the Hubble parameter.  In that case, the terms proportional to $\Psi_{\vec{k}}$ in Eq.~(\ref{PsieqnX}) sum to zero, and we are left with
\beq
\Psi^{\prime\prime}_{\vec{k}}(a)+\frac{1}{a}\left(\frac{7}{2}+\frac{3}{2}w\right)\Psi^\prime_{\vec{k}}(a) = 0.
\eeq
A constant value of $\Psi_{\vec{k}}$ is the only non-decaying solution to this equation.  Therefore, $\Psi$ is constant in the early universe, and we may set $\Psi^\prime$ equal to zero as an initial condition when numerically solving Eq.~(\ref{PsieqnX}).  Moreover, $\Psi_{\vec{k}}$ is always constant if $\omegaX=1$.

Now that we have expressions for $\Psi(a)$ and $\vec{v}$, the only remaining component of the dipole anisotropy is the SW effect.  Eq.~(\ref{SW}) only applies to universes that were initially radiation-dominated, so we need to derive the analogous expression for a universe that is initially $X$-dominated.   Given that the perturbations are adiabatic, the perturbation to the radiation density will be proportional to the density perturbation in the $X$ fluid:
\beq
\frac{\Delta T}{T}= \frac{1}{4}\delta_\gamma = \frac{1}{3(1+w)} \delta_X,
\eeq
where $\delta_\gamma$ and $\delta_X$ are the fractional density perturbations in the radiation and the  $X$ fluid, respectively, in conformal Newtonian gauge.   The superhorizon limit ($k\rightarrow0$) of the temporal Einstein equation, with $\dot\Psi=0$, implies that $\delta_X = -2\Psip$ at very early times.  We can then use the adiabatic condition to obtain the primordial temperature anisotropy:
\beq
\frac{\Delta T}{T}(\tau_\mathrm{p}) = -\frac{2}{3(1+w)}\Psip.
\label{initX}
\eeq
The Boltzmann equations for superhorizon perturbations still imply that $\Delta T/T = \Psi(\tau)$ plus a constant, and that constant is determined by Eq.~(\ref{initX}).  The final expression for the SW effect is 
\beq
 \frac{\Delta T}{T}(\taudec)+\Psi(\taudec) = \Psi(\taudec)\left[2-\frac{5+3w}{3(1+w)}\frac{\Psip}{\Psi(\taudec)}\right],
 \label{SWX}
 \eeq
which is equivalent to Eq.~(\ref{SW}) if the $X$ fluid is radiation ($w=1/3$).

We now have all the components necessary to numerically evaluate the observed CMB temperature anisotropy following the same procedure as described in Section \ref{sec:potential}.  The SW anisotropy ${\cal S}$ is given by Eq.~(\ref{SWX}).  The ISW anisotropy may be obtained by numerically solving Eq.~(\ref{PsieqnX}) and using that solution to evaluate $\scri0 -\scri1$.  Finally, Eq.~(\ref{veqnX}) gives the Dopper anisotropy $[{\cal V}(\tau_0)-{\cal V}(\taudec)]$.  Combining these terms, we find that $\delta_1=0$ for any value of $w\geq1/3$ and any value of $\omegaX$.   

In the case that $\omegaX=1$, the dipole cancellation is easy to see analytically.  Since $\Psi$ is constant, there is no ISW effect.  The Doppler anisotropy is given by 
\beq
{\cal V}(\tau_0)-{\cal V}(\taudec) = \frac{2}{3(1+w)H_0\xdec}\left[-1+\frac{H(\adec)}{H_0}\adec^{3w+2}\right].
\label{analX}
\eeq
Since the $X$-fluid is the only energy in the universe, the Hubble parameter is simply $H(a)=H_0 a^{-3(1+w)/2}$, and the comoving distance to the surface of last scattering is
\beq
H_0 \xdec = \frac{2}{1+3w} \left[1-\adec^{(1+3w)/2}\right].
\eeq
Inserting these expressions into Eq.~(\ref{analX}) yields
\beq
{\cal V}(\tau_0)-{\cal V}(\taudec) = -\frac{1+3w}{3(1+w)},
\eeq
which exactly cancels the SW anisotropy given by Eq.~(\ref{SWX}) with $\Psip =\Psi(\taudec)$.

\end{document}